\documentclass[11pt]{article}

\usepackage[english]{babel}
\usepackage[a4paper,nohead,twoside]{geometry}
\usepackage{amsmath}
\usepackage[mathscr]{euscript}
\usepackage{amsfonts}
\usepackage{amssymb}
\usepackage{amsthm}
\usepackage{mathtools}
\usepackage{graphicx}
\usepackage{color}
\usepackage{enumitem}
\usepackage{xspace}

\usepackage[bookmarks=true,
            bookmarksopen=true,colorlinks=false,breaklinks=true]{hyperref}

\numberwithin{equation}{section}

\newcommand \del 		\partial

\begin{document}

\title{General Relativity, Time, and Determinism}
\author{James Isenberg\footnote{Department of Mathematics and Institute of Theoretical Science, University of Oregon. \newline  isenberg@uoregon.edu}}

\maketitle

\begin{abstract}
Einstein's theory of general relativity models the physical universe using spacetimes which satisfy Einstein's gravitational field equations. To date, Einstein's theory has been enormously successful in modeling observed gravitational phenomena, both at the astrophysical and the cosmological levels.

The collection of spacetime solutions of Einstein's equations which have been effectively used for modeling the physical  universe is a very small subset of the full set of solutions. Among this larger set, there are many spacetimes in which strange phenomena related to time are present: There are solutions containing regions in which determinism and the predictability of experimental outcomes breaks down (the Taub-NUT spacetimes), and there others in which the breakdown of determinism occurs everywhere (the G\"odel universe).

Should the existence of these strange solutions lead us to question the usefulness of Einstein's theory in modeling physical phenomena? Should it instead lead us to seriously search for strange time phenomena in physics? Or should we simply treat these solutions as anomalous (if embarrassing) distractions which we can ignore?

In this essay, after introducing some basic ideas of special and general relativity and discussing 
what it means for a spacetime to be a solution of  Einstein's equations, we explore  the use of spacetime solutions 
for modeling astrophysical events and cosmology. We then examine some of the spacetime solutions in which determinism and causal relationships break down, we relate such phenomena to Penrose's ``Strong Cosmic Censorship Conjecture", and finally we discuss the questions noted above.

\end{abstract}




\section{Introduction}
\label{Intro}

Two of the most widely-known effects of relativity involve time. The ``Twin Paradox" describes the difference in elapsed time which a pair of twins will record if one of them travels very fast (close to the speed of light) relative to the other. The ``Gravitational Time Dilation" also describes a difference in elapsed time recorded by a pair of twins; in this case, between meetings, one of them lives in a much stronger gravitational field than the other.

Both of these effects are fairly straightforward predictions of relativistic mechanics. As such, they are necessarily present in any model of the universe which is consistent with the broad principles of special and general relativity. Not surprisingly, both of these effects have been directly observed in our universe. As well, although these relativistic  time measurement effects  were originally perceived as counter-intuitive and even disturbing (note the name ``Twin \emph{Paradox}"), physicists now are quite reconciled to their presence in our physical universe.

General relativity predicts the possibility of other strange effects related to time. Specifically, there are models of the cosmos consistent with the principles of general relativity in which observers can move cyclically in time. As well, there are models in which causal relationships  and determinism breakdown.  Should such possibilities be judged as realistic (and fascinating) predictions? Should they instead be judged as an indication that Einstein's theory of general relativity is seriously flawed?

To be able to discuss these two alternatives, and to possibly choose between them, it is important to understand what  general relativity is, and what its role is in modeling the physical universe. Although a full understanding of general relativity requires one to know some differential geometry and to know how to work with partial differential equations (PDEs), some of its key ideas can be understood without these mathematical tools. We present some of these key ideas in Section 2, including the notion of a spacetime solution of Einstein's equations. These ideas are discussed within the context of the  conceptual development of special and general relativity. In Section 3, we explore the use of spacetime solutions of Einstein's equations for modeling gravitational physics, noting the difference between the modeling of astrophysical events and the modeling of the full cosmos. We proceed in Section 4 to consider spacetime solutions in which causality and determinism break down.  We note the sense in which these effects are consistent with general relativity, and we describe the Strong Cosmic Censorship conjecture and the sense in which this conjecture argues that these effects are essentially irrelevant for modeling our universe. We make concluding remarks in Section 5. 

\section{Special Relativity and General Relativity}
\label{SR and GR}

The conceptual framework used by scientists to model the physical world at extreme scales changed profoundly during the first two decades of the Twentieth Century. The fairly intuitive ideas of Newtonian physics (developed during the late 1600s) work very well for human-scale physical experiments such as tossing balls and sending space probes to Pluto. However, to understand the interactions of subatomic particles, to study the dynamics of objects moving anywhere near the speed of light, and to predict what happens in the presence of extremely dense concentrations of mass and energy, it was found that radically new ways of thinking are needed. 

We do not explore here the new ideas---quantum mechanics and quantum field theory---that are needed to work with subatomic and elementary particle physics. Rather, we focus on special relativity and general relativity, which are needed to study the behavior of objects which move very fast, and objects which are very massive and concentrated.

Gedanken (thought) experiments heavily influenced Einstein's development of both special and general relativity. In the case of special relativity (SR), it was thinking about electromagnetism that was most influential. Einstein was driven to understand how the outcomes of gendaken experiments with moving magnets and moving conducting coils could be consistent with Maxwell's  theory of the electromagnetic field (published in 1865).  Einstein found that he could obtain this consistency only if (a) he treated the speed of light as an observer-independent physical constant, and (b) he dispensed with absolute measures of space and time, replacing them with locally determined (observer-dependent) measures of distances and time intervals. Both of these principles are completely at odds with the foundations of Newtonian theory, which presumes that all observers (regardless of their relative motion)  measure distances and time intervals identically; and presumes that if two observers measure  the motion of a light ray (or anything else), then their measurements must differ in accord with their relative motion.

Einstein based his formulation of special relativity on these two principles, together with the empirically-based idea that to make measurements, an observer must rely on a personal clock along  with a device for emitting and detecting directed light rays. Thus, to measure the length of a rod some distance away, an observer bounces light rays off each end of the rod and uses the clock-measured time interval between the return of the reflected rays to determine this length. Based on this measurement procedure, along with the assumption that the speed of light is identical for all observers, it is straightforward to calculate familiar SR phenomena such as the Lorentz length contraction, which relates the measured length of a rod as seen by two observers who are moving relative to each other. 

Special relativity provides scientists with a new way to think about measurements of time and space which fits beautifully with Maxwell's theory of electromagnetism. SR does this by prescribing how different (relatively moving) observers $\mathcal{O}_1$ and $\mathcal{O}_2$  measure different electric  and magnetic fields, and then showing that the resulting observations by $\mathcal{O}_1$ and $\mathcal{O}_2$ of these fields in the presence of moving magnets and conducting coils (as per one of Einstein's gedanken experiments) are each consistent with Maxwell's theory.  

Special relativity is \emph{not}, however,  consistent with Newton's theory of the gravitational field. This inconsistency is evident, since Newton's theory predicts that variations in the gravitational field which are generated by the motion of massive objects are manifest everywhere immediately, which contradicts  the SR principle that no signals can be transmitted faster than light-speed. As well, there are no transformations of the Newtonian gravitational fields such that the experimental measurements made by observers in different frames (with lengths and time intervals transforming according to the rules of SR) are each consistent with Newton's theory.

This inconsistency led Einstein to focus for the next ten years---from 1905 to 1915---on finding a new theory of the gravitational field.  Besides requiring this new theory to be consistent with the principles of special relativity---no signals traveling faster than the speed of light, and all measurements based on local considerations---Einstein believed it to be essential that his new theory of gravity incorporate the \emph{Equivalence Principle}. In its simplest form, the Equivalence Principle embodies the experimental fact that if a pair of bodies move in a fixed gravitational field---say, that of the earth---with no other forces present (e.g., no friction or electromagnetic forces), then their motion is identical, regardless of their masses or composition.  Newton's theory incorporates the Equivalence Principle in a somewhat ad hoc way: According to Newton, the acceleration $a$ of a given body is determined by the imposed force $F$ divided by the mass $m$ of that body, and if the force $F$ is gravitational, then $F$ is proportional to $m$. Hence for a body moving in a gravitational field the mass factor $m$ cancels, and consequently the acceleration induced by the gravitational force is independent of $m$. This works. However, Einstein believed it to be crucial for his new theory of gravity that the Equivalence Principle be built into the theory in a more essential way. 

The most striking feature of general relativity---the name Einstein chose to label his new theory of  gravity---is its introduction of \emph{curved geometry} into the setup it uses to model the universe and the motion of bodies contained in it. The idea is simple, but revolutionary: Instead of thinking about the universe as a flat, featureless, static  background stage in which bodies move in response to imposed forces, one thinks of it as a dynamic, curved space + time geometry in which the motion of bodies is determined by this geometry. 
More specifically, through each point in the spacetime and for each choice of a (local) velocity, the geometry determines a unique path. This path, which is called a \emph{geodesic}, is the free-fall path that a body with the prescribed initial velocity passing through the prescribed point will follow, \emph{regardless of its mass or composition}. In this way, the use of curved spacetimes allows general relativity to incorporate the Equivalence Principle in an essential way. 

It is important to note that the geometry we are discussing here characterizes the \emph{spacetime} as a unified entity, not just the space as something separate from time. One of the important innovations of special relativity is the unification of space and time into spacetime.  Doing this allows one to recognize the physical equivalence of different frames of reference related by the motion of one of the frames relative to the other. It also provides a very useful way of visualizing such things as how different observers perceive the simultaneity of spatially-separated events in different ways. The combination of three-dimensional space and time into four-dimensional spacetime  is a key feature of general relativity as well as special relativity. Indeed, spacetimes with  specified curved geometries (with their corresponding arrays of geodesic paths) are the fundamental objects which are studied in general relativity, and are used to model gravitational physics. 

What determines the curvature of a spacetime used to model the universe (or portions of it), according to Einstein's theory of general relativity? It is often stated that while the spacetime curvature determines how matter moves, the matter determines how the spacetime curves. This is only partially true. The key to understanding this is the Einstein gravitational field equation, which takes the form $G_{\mu \nu} = \kappa T_{\mu \nu}$. The object on the left hand side of this equation represents the spacetime curvature at any given point in spacetime (it is known as the Einstein curvature tensor field). The important thing to note about $G_{\mu \nu}$ is that it controls only a portion of the spacetime curvature---roughly half of it at each point in the spacetime. Even if the Einstein tensor is zero everywhere, the spacetime can be very curved. As for what appears on the right hand side of this equation, besides the constant factor $\kappa$ (which depends on the speed of the light and the universal gravitational constant from Newton's theory) one has the stress energy tensor field $T_{\mu \nu}$. This object represents the localized mass density and momentum density and angular momentum density of matter  and non-gravitational fields (including, e.g., the electromagnetic field) at each point in the spacetime. So, while the matter and the fields and the curvature in a general relativistic spacetime must satisfy Einstein's equation $G_{\mu \nu} = \kappa T_{\mu \nu}$, this relation by no means  implies that matter determines curvature. If it did, then general relativity would predict that gravitational radiation does not exist; this of course would be inconsistent with the recent LIGO observations of gravitational waves. 

\section{General Relativity and the Modeling of our Universe}
\label{Modeling}

General relativity is used to model gravitational effects at three very different scales: a) Near-earth phenomena, such as the gravitational effects of the earth on GPS signals; b) astrophysical events, such as the collision of a pair of black holes; and c) cosmological features, such as the production of the cosmic microwave background by the Big Bang. There is an important conceptual difference between the modeling used for cosmological studies as opposed to that used for near-earth and astrophysical phenomena. Regarding  the latter two cases, there is a very wide variety of different physical systems of interest which are expected to exist somewhere in the universe, and one studies a particular one  by finding a spacetime solution of the Einstein equations which corresponds to that system. Note that such a spacetime solution is not expected to describe the entire universe; it is designed to model a local (relatively isolated) physical system in a very small portion of the universe. Each such spacetime solution may well be physically relevant, describing physical phenomena in widely separated portions of our universe.

In contrast, in modeling cosmological phenomena using general relativity, one works with solutions of Einstein's equations which are supposed to represent the entire universe. Since we live in just one universe\footnote{In this essay, we ignore  the possibility that we live in a ``multiverse", with regions that will never be observable.}, in principle only one spacetime solution is needed for cosmology, and only one is completely accurate. The catch is, we don't know enough about our universe to narrow down which spacetime solution to use for cosmological modeling. Consequently, in doing cosmology, we are led to consider many solutions, hoping that such a wide-ranging study can be useful for learning about the cosmology of our particular unique universe.

To illustrate the difference between these two types of modeling, it is useful to discuss  an example of each kind:    i) modeling the gravitational radiation produced by the collision of two black holes, and ii) modeling the universe immediately after the Big Bang. 

One of the most exciting developments in physics in the 21st Century thus far  is the direct detection of gravitational radiation for the very first time, by LIGO (the Laser Interferometry Gravitational Observatory, located both in Washington state and in Louisiana) \cite{LIGO}. Gravitational radiation is effectively ripples in the curvature in the  spacetime, generated primarily by accelerating concentrations of matter. Though ubiquitous, such radiation is generally extremely weak, and consequently very difficult to detect. To enable it to be detected, as well as to be interpreted, it is crucial to be able to accurately model the gravitational radiation which is expected to be produced by very strong sources such as a pair of colliding black holes. Based on its remarkable success in modeling other gravitational effects such as the observed changes in the light signals emitted from the Hulse-Taylor binary pulsar \cite{Hulse-Taylor}, general relativity is used to carry out this modeling. 

Conceptually, the modeling of the  collision of a pair of two black holes is simple: First, one chooses the distinguishing parameters of the collision: the masses and the spins of each of the black holes, their initial separation and initial relative velocity, and the relative directions of the spins and velocities. This choice picks out the one particular black-hole collision of interest. Next, in accord with the choice of these parameters, one designates  the initial data for the collision. This consists of the snapshot initial geometry and the initial rate of change of the geometry. Besides matching the choice of the parameters of the particular collision being modeled, the designation of the initial data must also satisfy a set of initial-data-constraint equations; corresponding to four of the ten Einstein gravitational field equations, these are analogous to the Maxwell constraint equations $\nabla \cdot B=0$ and $\nabla \cdot E= 4\pi \rho_{charge}$ which the electric and magnetic fields must satisfy. After the designation of the initial data is made, one uses the remaining six of the Einstein gravitational field equations to evolve the geometry into a spacetime which satisfies the full system $G_{\mu \nu} = \kappa T_{\mu \nu}$ everywhere. From this spacetime, with a bit of straightforward work, one deduces such things as how long it takes for the black holes to collide and merge, how much gravitational radiation is emitted, and what the particular profile of the emitted radiation is. 

As noted above, modeling black-hole collisions in the way just described is crucial to the success of LIGO in detecting and in analyzing gravitational radiation. While it took well over thirty years to work out the details of how to carry out  this sort of modeling numerically, the process is now to a large extent routine.\footnote{Routine, but very time consuming: Numerical runs can take hundreds of hours.} It provides a wonderful example of the role that solutions of Einstein's equations can play in astrophysics.

At the time of this writing, only one black-hole collision has been detected (and confirmed) by LIGO. One expects, however, that many more such collisions will soon be detected,. Consequently it is very likely that spacetime solutions of Einstein's equations corresponding to the full range of the parameter space of black-hole collisions will each be useful in astrophysical modeling. 

The use of general relativity to construct models of the full cosmos is very different from its use in modeling astrophysical events like black hole collisions. We live in a unique universe, so in principle just one solution of Einstein's equations is useful for modeling it in detail. However, as noted above, since we know so little about the full cosmos, we can not hope to know which is the specific spacetime solution which most accurately models our universe. Consequently we are led to construct a wide variety of solutions, not knowing which may be  useful and which are not. 

For example, say we wish to consider the question of how it is that the cosmic microwave background (CMB) radiation \cite{CMB}, which is believed to be a relic of the Big Bang over thirteen billion years ago, is observed to be very  nearly the same in all directions, yet not exactly the same in all directions (i.e., nearly isotropic, but not  exactly isotropic). One way to explore this question is to consider all spacetime solutions which evolve from a Big Bang and are broadly consistent with other features of the universe such as its age and apparent matter content, and then try to show that some large portion of these solutions produce nearly isotropic CMB radiation. It is of course impossible to construct all such solutions, even through numerical simulations, One can, however, focus on a subset of them---characterized, for example, by some symmetry---and examine the generic behavior of solutions in this subset. This approach played an important role in convincing many that Einstein's field equations with standard matter fields are not enough to model our universe. Rather, it appears some mechanism for producing inflation in such models is likely needed \cite{inflation}.  

It is important to note that there are many spacetime solutions of the Einstein equations which are clearly \emph{not} expected to be of any direct use for modeling our universe and the gravitational phenomena which may occur in it. For example, cosmological solutions which are static, or which go from a big bang to a big crunch in a very short time, are useless for modeling. This feature distinguishes general relativity from other classical field theories such as Maxwell's theory of electromagnetism. In the case of Maxwell's theory, one can plausibly argue that essentially any solution might serve  to model electromagnetic phenomena somewhere in the universe. Indeed, if one focuses on solutions of Einstein's equations which are designed to model localized astrophysical events, then the same argument for the potential usefulness of all such solutions might be made. The feature of general relativity that leads to clear disqualification of some  solutions is its service for modeling the entire cosmos, and not just localized phenomena.

\section{Causality, Determinism, and Solutions of Einstein's Equations}
\label{Causality}

Deeply ingrained in our concept of how science---at least, physics---works is the idea of the \emph{deterministic experiment}: One specifies the initial state of the system---say, the initial position and velocity of a ball near the surface of the earth---and then the system is compelled by the ``laws of physics" to evolve in a unique, prescribed way.  For example, for the ball near the surface of the earth, Newton's theory prescribes the acceleration of the ball, and hence (with the initial position and velocity specified) it determines a unique subsequent path for the ball.\footnote{Einstein's theory prescribes essentially the same path.} 

While there is nothing that tells us that physics \emph{has} to work this way, it is a measure of the success of our science that we have been able to find theories which---at least within the realm of physical phenomena for which quantum theory is not needed---tell us exactly what ``initial data" it is sufficient for us to know for a given system so that we can use  the theory to calculate the future evolution of that system accurately. Notably, this works perfectly for electromagnetic phenomena as modeled by Maxwell's theory in the context of special relativity: Presuming that there is no charged matter around\footnote{The presence of charges does not change this, so long as a theory modeling the behavior of charges is prescribed along with Maxwell's theory.}, if we choose a global inertial frame\footnote{Special relativity allows this.}  and if we know the electric and magnetic fields everywhere in space at a given moment of time (relative to this chosen frame), then Maxwell's equations determine these fields everywhere to the future as well as to the past. 

If, instead of wanting to determine the electromagnetic fields everywhere for all time, we only seek to determine those fields at some particular point in space\footnote{Here and below, we use single latin letters such as $``x"$ to label spatial points, even though in terms of coordinates, one needs three letters to label such points.}   $x$ and at some particular time $T$ in the future, do we need to know what the fields are now  \emph{everywhere} in space?  Presuming for the moment that we are considering this problem in the flat spacetime of special relativity (known as the ``Minkowski spacetime"), then in fact we only need to know the values of the fields ``now" (which we label  as time $t=0$) in a particular region. This region, which we label $\mathcal {P}_{[x,-T]}$, consists of all those points $y$ such that a light ray or a material object might travel from $y$ at time $t=0$ to $x$ at time $t=T$ (a path in spacetime which might in principle be traversed by either a light ray or  by a material object going slower than the speed of light is called a \emph{causal path}).  The values of the fields at time $t=0$ (now) which are outside $\mathcal {P}_{[x,-T]}$ are completely irrelevant to determining the fields at $(x,T)$, because no signal from this outside region can travel fast enough (faster than the speed of light) to get to the point $x$ at time $T$. 

Correspondingly, still restricting ourselves to the physics of special relativity, we see that if we choose a region in space $\Sigma$ at a time $t=0$, then there is a collection of spacetime points $(z,t)$ to the future of $(\Sigma, t=0)$ such that every causal path which hits one of the points $(z,t)$ \emph{must} pass through $(\Sigma, t=0)$, and also  such that no causal path which hits  these points $(z,t)$ may pass through any points outside of $(\Sigma, t=0)$ at time $t=0$. This collection of points is called the future  \emph{domain of dependence}\footnote{There is a corresponding past domain of dependence $\mathcal{D}^-(\Sigma_{t=0})$, defined analogously.} of $\Sigma$, and is labeled $\mathcal{D}^+(\Sigma_{t=0})$. It follows from special relativity and Maxwell's theory  that the electromagnetic fields in $\mathcal{D}^+(\Sigma_{t=0})$ are completely determined by the initial data of the fields on $(\Sigma, t=0)$.

The ability to determine the future evolution of physical systems from initial data, and in particular the ability to do this in a localized way as described above via such constructs as $\mathcal {P}_{[x,-T]}$ and $\mathcal{D}^+(\Sigma_t)$ is a key feature of special relativity. The  language used to affirm this feature is that Minkowski spacetime is \emph{globally hyperbolic} and does not violate \emph{causality}. 

Does this same sort of thing work with physical phenomena for which general relativity is needed? One of the fascinating features of general relativity is that for certain classes of spacetime solutions it does, while for others it does not. 

It is easy to see that there are spacetimes which satisfy Einstein's equations of general relativity, yet fail to be globally hyperbolic. To construct an example (which we label ``identified-Minkowski spacetime"), we take the standard flat Minkowski spacetime with a standard set of coordinates $(x,t)$, we throw out all of the spacetime with $t>1$ or with $t<0$, and then for each choice of the spatial coordinates $(x)$, we identify the spacetime points $(x,0)$ and $(x,1)$. Since the curvature is zero everywhere and since there is no matter around anywhere, the equations $G_{\mu \nu}=T_{\mu \nu} =0$ are certainly satisfied everywhere. Furthermore, despite the somewhat bizarre identification of spacetime points which has  been made in constructing this spacetime solution, it does not violate any explicit rules for general relativity. Yet, with a bit of thought we see that if we choose a point, say $(x, 3/4)$, in the spacetime and then seek to identify the region $\mathcal {P}_{[x,-1/2]}$ 
as a subset of the spacetime with $t=1/4$, we are forced to include \emph{all} spatial points at that value of $t$. This is true because for \emph{any} point $(y, 1/4)$, there is a causal path in this bizarre spacetime which connects $(y, 1/4)$ and $(x, 3/4)$ (it may have to pass through the  $t=0 \leftrightarrow  t=1$ identification several times). Similar considerations show that for any choice of a spatial region $\Sigma$ at any time $t_0$, the domain of dependence 
$\mathcal{D}^+(\Sigma_{t_0})$ in this spacetime is empty. It follows from these strange features that deterministic experiments do not make sense in this spacetime, since there are no sets of initial conditions for the electromagnetic field (or for any other field) in some region $\Sigma$ at some time $t_0$ which determine the behavior of that field anywhere into the future. We also see that this spacetime contains causal paths which close on themselves; hence the notions of ``future" and ``past" lose their meaning in this spacetime.

To avoid determinism and causality problems of the sort just described, one might require that spacetime solutions of Einstein's equations have the topology of $\mathbb{R}^4$. One of the exciting features of general relativity from its very beginnings \cite{Einstein}, however, has been its opening up of the possibility of working with spacetimes with topologies more general than $\mathbb{R}^4$. Indeed, until fairly recently, many cosmologists believed that the most useful spacetimes for modeling our universe were likely to be spatially closed, with a spacetime topology of the form $S^3 \times \mathbb {R}$ (where $S^3$ is the three-dimensional sphere).

To allow spacetimes with interesting topology, while disallowing fairly contrived spacetimes such as identified-Minkowski, one might decide to include in the stipulations of the theory of general relativity the requirement that the spacetime topology be $\Xi \times \mathbb{R}$, where $\Xi $ could be any three-dimensional manifold\footnote{ An n-dimensional  manifold is a space which locally looks like $\mathbb{R}^n$. The three-dimensional sphere is an example.}. One might then ask if this restriction to the theory prevents not just identified-Minkowski spacetime, but also throws out any spacetime solution with causality and global hyperbolicity problems. 

In fact it does not. To illustrate this, we discuss two very different archetypal examples here: the Taub-NUT spacetime and the G\"odel spacetime. The Taub-NUT spacetime \cite{Misner} which has the topology $S^3 \times \mathbb{R}$, is a solution of the Einstein equations with no matter present. It contains a spacetime region (the ``Taub region") in which the spacetime is fully deterministic:  Within the Taub region, the spacetime has no closed (or almost closed) causal paths and therefore does not violate causality. As well, in the Taub region, domains of dependence can be localized as in special relativity, and since the Taub region lies inside the union of the future and the past domains of dependence of any $S^3$ hypersurface labeled with  a fixed choice of Taub-NUT time, the gravitational field of the Taub region is determined by gravitational initial data on such a hypersurface. The same would hold for other fields (such as electromagnetic fields) on a Taub-NUT spacetime background. The Taub region is thus labeled as ``globally hyperbolic". 

The Taub region is bounded by a particular $S^3$ hypersurface (called a ``Cauchy horizon");  passing beyond it into the NUT region, one finds that causality, global hyperbolically, and determinism all break down. There are closed causal paths, domains of dependence become empty, and the evolution of fields is not determined by specified sets of initial data. Indeed there are multiple possible NUT regions which can be smoothly attached (as solutions) to the Taub region, thus constituting multiple possible ``futures" of the Taub region. 

By contrast with the Taub-NUT spacetime, the G\"odel spacetime \cite{Godel} has no region in which it is causal or globally hyperbolic or deterministic in any sense. A solution of the Einstein equations with ``dust"-type  matter (with non-vanishing vorticity) and with a cosmological constant, the G\"odel spacetime is topologically simple: $\mathbb{R}^4$. However, through every one of its points, there are many closed causal paths. Past and future make little sense, and deterministic experiments cannot be carried out in a G\"odel spacetime. 

Are the G\"odel and Taub-NUT spacetimes anomalous examples, which should best be hidden in the closet and ignored? While it is not at all clear if the G\"odel spacetime in any sense exemplifies a class of solutions with similar properties, in fact there is a wide class of solutions with properties very similar to the Taub-NUT solution \cite{Moncrief}. Notably, these ``generalized Taub-NUT solutions" are considerably less specialized than the Taub-NUT spacetime itself, since they are much less symmetric\footnote{The Taub-NUT spacetime is invariant under the action of a three-dimensional isometric group, while there are known generalized Taub-NUT solutions with only a one dimensional isometric group.}.  There is an infinite dimensional family of them, each one containing a globally hyperbolic region in which determinism holds, and each one extendible (in multiple ways) across a Cauchy horizon into a non globally hyperbolic region with closed causal paths. 

Back in the 1960s, after he and Stephen Hawking had proven their celebrated spacetime incompleteness theorems, Roger Penrose \cite{Penrose} proposed a pair of conjectures which have become known as Weak Cosmic Censorship (WCC) and  Strong Cosmic Censorship (SCC). The spacetime incompleteness theorems, often called  ``singularity theorems", show that if a spacetime solution satisfies a fairly general set of hypotheses, then it necessarily contains causal paths which are forced to stop within a finite period of (local) time\footnote{A spacetime containing such paths is called `` geodesically incomplete".}. The reason such paths are forced to stop could be because the spacetime contains a region in which the curvature blows up; the presence of a Cauchy horizon could cause this as well. Penrose's Strong Cosmic Censorship conjecture proposes that in almost all such cases, it is curvature blowup rather the presence of a Cauchy horizon which causes the demise of causal paths\footnote{Weak Cosmic Censorship is essentially unrelated to Strong Cosmic Censorship. WCC conjectures that in essentially all astrophysical-type spacetime solutions which contain unbounded curvature, distant observers cannot see signals from the  region in which this occurs; that is, such regions must be contained inside black holes.}.

If Strong Cosmic Censorship is true, then spacetimes (such as the generalized Taub-NUT) which contain Cauchy horizons (and the other attendant difficulties with determinism) are a very small subset of the collection of all solutions of Einstein's equations. Is SCC in fact true? 

Although Strong Cosmic Censorship has often over the past fifty years been cited as one of the major questions in the mathematical study of general relativity, it is far from clear whether SCC is true or not. Model versions of the conjecture have been proven in small families of solutions \cite{SCC}; there are also recent results which suggest that SCC is not likely to hold in its strongest form. The verity of  the conjecture remains a wide open question. 

\section{Conclusion}
We know two important things about general relativity: 1) Observation and experiment have shown that it is extremely effective for modeling gravitational physics, from the scale of the solar system to the scale of astrophysical phenomena. 2) It includes among its spacetime solutions a number of them in which causality and determinism break down, at least in certain regions. 

Based on these two facts, one might be led to believe one of the following assertions:
A) We should expect to find causality violations and the failure of determinism somewhere in our universe.\\
B) Since we have not detected any breakdown of causality or determinism in our universe, and since they are so fundamental to our way of doing science, general relativity must be seriously flawed as a physical theory to be used for modeling our universe.\\
C) While general relativity includes among its array of solutions some in which causality and determinism fail, these solutions are irrelevant for modeling physics, and can be more or less ignored.

All three of these statements are consistent with what we know about general relativity. Which of them makes the most sense scientifically? 

Since we know already that there are many solutions of Einstein's equations which are of no use for modeling physics (see Section 3), and since we in fact live in just one universe, it is hard to support statement B. Whether or not one chooses to invoke a selection principle (e.g., restrictions on topology, restriction to globally hyperbolic solutions, etc.) to determine which spacetime solutions might be used for modeling physics, it is straightforward to focus on certain solutions and ignore others in carrying out modeling. That this must be done in using general relativity does not, I believe, harm the usefulness of the theory. One might argue that this makes it more difficult to believe that Einstein's theory is \emph{right}. However, we already know that Einstein's theory cannot be used to model systems in which quantum ideas are needed, so this argument is a red herring. As well, it is very likely that \emph{no theory} that we know now (or will ever know?) will prove to be \emph{right} in the sense that it explains and models phenomena at all scales. 

Regarding statement A, it of course makes sense on a grand scientific scale to search for situations (as ``predicted" by solutions of Einstein's equations) in which phenomena corresponding to the breakdown of causality and determinism might be detected. The predictive  solutions could be used as a guide to finding such phenomena. On the other hand, one might argue that while it would be fascinating to find such phenomena, they  are likely to be very difficult to find; hence such searches should be of very low priority. 

Statement C is likely the most practical scientific position to take. A wide range of spacetime solutions of Einstein's equations are extremely effective for modeling and predicting gravitational physics. The existence of other, fairly strange, solutions is interesting, but perhaps irrelevant scientifically. 

Perhaps; we shall see.

\section*{Acknowledgements}
This work was partially supported by NSF grant  PHY-1306441 at the University of Oregon. I thank the Mathematical Sciences Research Institute  for hospitality during the course of the writing of portions of this review.

\bibliography{bibliography}

\end{document}